\documentclass[12pt]{article}
\usepackage{amsmath}
\usepackage{amssymb}
\usepackage{latexsym}

\begin{document}

\baselineskip 6mm
\renewcommand{\thefootnote}{\fnsymbol{footnote}}

\newcommand{\nc}{\newcommand}
\newcommand{\rnc}{\renewcommand}


\rnc{\baselinestretch}{1.24}    
\setlength{\jot}{6pt}       
\rnc{\arraystretch}{1.24}   




\nc{\be}{\begin{equation}}

\nc{\ee}{\end{equation}}

\nc{\bea}{\begin{eqnarray}}

\nc{\eea}{\end{eqnarray}}

\nc{\ben}{\begin{eqnarray*}}

\nc{\een}{\end{eqnarray*}}

\nc{\xx}{\nonumber\\}

\nc{\ct}{\cite}

\nc{\la}{\label}

\nc{\eq}[1]{(\ref{#1})}

\nc{\newcaption}[1]{\centerline{\parbox{6in}{\caption{#1}}}}

\nc{\fig}[3]{

\begin{figure}
\centerline{\epsfxsize=#1\epsfbox{#2.eps}}
\newcaption{#3. \label{#2}}
\end{figure}
}


\def\IR{{\hbox{{\rm I}\kern-.2em\hbox{\rm R}}}}
\def\IB{{\hbox{{\rm I}\kern-.2em\hbox{\rm B}}}}
\def\IN{{\hbox{{\rm I}\kern-.2em\hbox{\rm N}}}}
\def\IC{\,\,{\hbox{{\rm I}\kern-.59em\hbox{\bf C}}}}
\def\IZ{{\hbox{{\rm Z}\kern-.4em\hbox{\rm Z}}}}
\def\IP{{\hbox{{\rm I}\kern-.2em\hbox{\rm P}}}}
\def\IH{{\hbox{{\rm I}\kern-.4em\hbox{\rm H}}}}
\def\ID{{\hbox{{\rm I}\kern-.2em\hbox{\rm D}}}}


\def\Tr{{\rm Tr}\,}
\def\det{{\rm det}}
\def\const{{\rm const}}


\def\vare{\varepsilon}
\def\barz{\bar{z}}
\def\barw{\bar{w}}

\begin{titlepage}
\hfill\parbox{5cm}
{DAMTP-2005-43 \\ \\ hep-th/0504158}\\
\vspace{25mm}
\begin{center}
{\Large {\bf Notes on non-extremal, charged, rotating black holes in
minimal $D=5$ gauged supergravity} }

\vspace{15mm}
Hari K. Kunduri \footnote{H.K.Kunduri@damtp.cam.ac.uk} and James
Lucietti\footnote{J.Lucietti@damtp.cam.ac.uk}
\\[3mm]
{\it  DAMTP, Centre for Mathematical Sciences,\\
      University of Cambridge,
      Wilberforce Rd.,\\
      Cambridge CB3 0WA, UK\\}

\end{center}

\thispagestyle{empty}

\vskip2cm


\centerline{\bf Abstract} \vskip 4mm \centerline{
\parbox[t]{15cm}{\small
\noindent We consider the non-extremal, charged, rotating black hole
solution of five dimensional minimal gauged supergravity of Cvetic,
Lu and Pope [Phys.\ Lett.\ B {\bf 598} (2004) 273]. We compute the Ashtekar-Magnon-Das mass and show it
agrees with the thermodynamic mass. We find a reducible Killing
tensor and integrate the geodesic equation explicitly. We also
compute the Euclidean action of the black hole and show it satisfies
the quantum statistical relation. Further we present a Smarr relation.
We end with a discussion of applications to string theory. }}

\vspace{2cm}

\end{titlepage}

\section{Introduction}
There has recently been renewed interest in solutions of five
dimensional gravity and supergravities, both in the ungauged and
gauged case. The bosonic sector of minimal supergravity is
Einstein-Maxwell with a Chern-Simons term in the former case. In the
latter case a negative cosmological constant is present.

Rotating higher dimensional black holes were first discussed in the
pure vacuum case in~\cite{MyersPerry}. However as is well known, it
is very difficult to find black holes with both rotation and charge
especially in the non-extremal case. Nonetheless significant
progress in the ungauged case has been made by considering
compactifications of higher dimensional theories and various
solution generating techniques~\cite{CveticYoum}. Analogous progress has been made
recently in the context of black rings. The original vacuum
solutions were found in~\cite{vacring}, and charged generalisations
followed~\cite{Elvang,EMDKL,dipolering}. Although these results
imply there exist black holes and black rings with the same
asymptotic charges, it seems in the extremal case, the
supersymmetric black ring~\cite{susyring} and BMPV black
hole~\cite{BMPV} do not have overlapping charges. The task of
constructing non-extremal black ring solutions has been initiated
in~\cite{nextremalring}.

In the gauged case, the search for charged, rotating black
holes has proven more difficult; however, due to the obvious
implications for the $AdS_{5}/CFT_{4}$ correspondence, it has attracted great
interest.  The uncharged solution with general rotation was found
in~\cite{HHT}; the five dimensional Reissner-N\"ordstrom-AdS
solutions were subsequently considered in~\cite{CEJM}. From the
context of string theory, these correspond to internal rotations in
the $AdS_5$ spacetime and black holes carrying appropriate $U(1)$ charge
in the $S^5$ Kaluza-Klein compactification respectively. A rotating
supersymmetric solution was found in~\cite{SaKl}, though it contains
naked closed timelike curves. Finally, using the
form of the general solution of the gauged theory found
in~\cite{GaGu}, further pathology-free examples were found and it was shown that, contrary to the minimal case, these
black holes \emph{had} to rotate~\cite{susyadsbh}.  Remarkably, a
non-extremal solution of gauged supergravity has been recently  found in~\cite{CLP}
with equal angular momenta parameters which contains the previous
supersymmetric solutions as particular limits.

The authors of~\cite{GPP}, motivated by the derivation of the
general Kerr-(anti) de Sitter black holes in all
dimensions~\cite{GLPP}, investigated the thermodynamic properties of
these solutions. In particular they showed how one must properly
define the mass and angular momenta in order to ensure the first law
of thermodynamics holds. They also used standard background methods
to compute the partition function rather than boundary counter-term
methods, pointing out inconsistencies in the previous literature.
The solution of~\cite{CLP} is thus interesting to study in this
context because it carries electric charge in addition to rotation
in two orthogonal planes. Analogous results have been obtained for
the four dimensional Kerr-Newman AdS black hole in~\cite{CCK}.
Properties of rotating, charged black holes in gauged supergravities
in four, five, and seven dimensions have also been examined recently
in~\cite{CGLP}.

In this article we will examine properties of this solution. In
section two we review the solution and how it is parameterized and
outline some basic characteristics. We follow this by looking at the
conserved charges (electric charge, momentum, and mass). The mass is
computed via the conformal boundary approach of Ashtekar, Magnon and
Das~~\cite{AD} and is shown to agree with the so called
`thermodynamic mass' found in~\cite{Ross} which was constructed to
satisfy the first law of black hole mechanics. In section four, we
consider symmetries of the spacetime and identify a reducible Killing
tensor which allows one to separate the Hamilton-Jacobi equation
explicitly. Next we turn to the computation of the Euclidean action
of the grand canonical ensemble. We use standard methods and not the
counter-term techniques. We show that our action and charges satisfy
the quantum statistical relation and are consistent with the first
law. We also give a Smarr relation. Finally we consider applications
to string theory and holography.

\section{Non-extremal charged, rotating black hole}
The bosonic action for $D=5$ minimal gauged supergravity is \bea \label{action} S = \frac{1}{16
  \pi} \int \left(\sqrt{|g|} ( R-12\lambda -F^2) - \frac{2}{3\sqrt{3}}
\eta^{\mu\nu\rho\sigma \delta} F_{\mu\nu}F_{\rho\sigma} A_{\delta}
\right) \eea where $\eta^{\mu\nu\rho\sigma\delta}$ is the
alternating symbol and $F=dA$.   The equations of motion that follow
are: \bea R_{\mu\nu} = 2 \left( F_{\mu
\rho}F_{\nu}^{\phantom{\nu}\rho}
- \frac{1}{6} F^2 g_{\mu\nu} \right) + 4\lambda g_{\mu\nu} \\
\nabla_{\mu}F^{\mu \nu} = \frac{1}{2\sqrt{3} \sqrt{|g|}} \eta^{\nu
\alpha\beta\gamma \delta} F_{\alpha\beta}F_{\gamma\delta}. \eea A
solution to this theory was found by Cvetic, Lu and Pope~\cite{CLP}
representing a black hole with spherical horizon topology
parameterised by its mass, electric charge and one independent
angular momentum. The solution they presented depends on an extra
parameter. However subsequently it was shown in~\cite{Ross}
that by performing a coordinate transformation $\psi \to \psi +
(\const)t$ one may remove this additional hair which is thus
unphysical. It is this simplified version of the metric we shall
work with. It is most concisely written in terms of left invariant
one forms $\sigma_i$ on $SU(2)$ which satisfy $d\sigma_1 = \sigma_2
\wedge \sigma_3$ and cyclic permutations thereof. Explicitly: \bea
\label{solution} ds^2 &=& -\frac{r^2W(r)}{4b(r)^2}dt^2
+\frac{dr^2}{W(r)} + \frac{r^2}{4}( \sigma_1^2 +\sigma_2^2) +
b(r)^2( \sigma_3 + f(r)dt)^2
\\  A &=& \frac{\sqrt{3}}{2} \frac{q}{r^2} \left(dt- \frac{1}{2}j \sigma_3
\right) \eea where the functions $W$, $b$ and $f$ are defined in the
Appendix. The event horizon is located at the largest real root of
$W(r)$, say $r=r_+$. Note that the absence of closed timelike curves
constrains the parameters somewhat as has already been discussed
by~\cite{CLP, Ross}. These black holes are in general
non-supersymmetric though in particular limits they reduce to the
Klemm-Sabra black hole ($p=0$)~\cite{SaKl} and the causally sound
solution of Gutowski and Reall~\cite{susyadsbh}. The reader is
referred to~\cite{Ross} for the details.

\section{Asymptotic charges}
In this section we will compute the mass, angular momentum and
charge of the black hole solution above. Since it is not
asymptotically flat, the calculation of mass is non-trivial. We want
to compare the gravitational mass to the `thermodynamic' mass
presented in~\cite{Ross}.
\par \noindent
The electric charge is defined as \be Q= \frac{1}{4\pi} \int_{S^3}
\left(
* F - \frac{2}{\sqrt{3}} A \wedge F \right) \ee
where $S^3$ is the $3$-sphere at infinity ($r \to \infty$, $t
=\const$), and the Chern-Simons term is required to ensure $Q$ is a
conserved quantity. For the black hole in question
 one finds that the Chern-Simons term does not contribute and the result is\footnote{In~\cite{Ross} it is stated that $Q=q$.
 Presumably they used a different normalisation.}:
 \be Q= \frac{\sqrt{3}}{2}\pi q .\ee The angular momentum may be
obtained via Komar\footnote{We define $J[m]=\frac{1}{16\pi}
\int_{S^3} *dm$. For the case at hand $m=2\partial_{\psi}$.}
integrals and for the Killing vector $2\partial_{\psi}$ is given
by~\cite{Ross} \be J= \frac{\pi}{4}j(2p-q). \ee Finally the mass may
be computed in a number of ways. Here we will consider the conformal
mass of Ashtekar, Magnon, and Das which was successfully used
in~\cite{GPP} to compute the energy of the uncharged Kerr-(A)dS
spacetimes in all dimensions. By `successfully' we mean that these
masses coincide with the thermodynamic mass.
\par
It is convenient to set $\lambda = -\frac{1}{\ell^2}$ to conform to
standard conventions.  Define the conformally rescaled metric
$\bar{g}_{ab} = \Omega^2g_{ab}$ where $\Omega = \frac{\ell}{r}$. As
$r \rightarrow \infty$ one sees the boundary of the conformally
rescaled metric is the Einstein static universe with metric
\begin{equation}\label{ESU}
d\bar{s}^2 = -dt^2 + \ell^2d\Omega_{3}^2
\end{equation} which may be thought of as the conformal
compactification of four dimensional Minkowski spacetime. As in the
uncharged case, the conformal boundary of the black hole is
$\mathbb{R} \times S^3$. Let $C^{\alpha}_{\phantom{a} \beta \gamma
\delta}$ be the Weyl tensor of either metric above. Setting $n = d
\Omega$, the electric part of the Weyl tensor is
\begin{equation}
\bar{\mathcal{E}}^{\mu}_{\phantom{a} \nu} =
\frac{\ell^2}{\Omega^2}\bar{g}^{\alpha\gamma}\bar{g}^{\beta\sigma}n_{\gamma}n_{\sigma}C^{\mu}_{\phantom{a}
\alpha \nu \beta}.
\end{equation}
The conserved quantity associated
with the Killing vector $K$ is
\begin{equation}\label{ADM}
Q[K] = \frac{\ell}{16\pi} \int_{\Sigma}
\bar{\mathcal{E}}^{\mu}_{\phantom{a} \nu}K^{\nu}d \Sigma_{\mu}.
\end{equation} Here the integral is taken over the $S^3$ at infinity
  with timelike normal $dt$. To compute the mass we take $K$ to be the
  timelike Killing vector $\emph{non-rotating}$ at infinity. From the
  leading order term of the relevant component of the Weyl tensor
\begin{equation}
C^{t}_{\phantom{a} r t r} = \frac{\ell^2(6p - 6q - 2\lambda
j^{2}p)}{r^6} + O(r^{-8})
\end{equation} it follows immediately $\bar{\mathcal{E}}^{t}_{\phantom{a} t} =
\frac{6p-6q-2\lambda j ^{2} p}{\ell^4}$. The metric on the conformal
boundary on constant time slices is simply that of an $S^3$ of radius
$\ell$. Inserting this into~(\ref{ADM}) yields
\begin{equation}\label{ADMmass}
Q[\partial_t] = \frac{\pi}{4}(3p -3q - \lambda j^2p)
\end{equation} which, as we indicate in a later section, is in
agreement with the thermodynamic mass. A further
quantity of physical interest is the gyromagnetic ratio of these
charged, rotating AdS black holes. For completeness we have computed
this: \be g= \frac{3p-3q-\lambda j^2p}{2p-q} \ee where $\mu_{ij} = g
\left( \frac{Q}{2M} \right) J_{ij}$, $\mu_{ij}$ are the dipole
moments and $J_{ij}$ are the angular momenta in Cartesian
coordinates. In the asymptotically flat case $ 3/2 \leq g \leq 3$
where the upper inequality is saturated if and only if $p=0$ which is the BPS
condition.

\section{Symmetries and geodesics}
The isometry group of the black hole is $\mathbb{R} \times SU(2)
\times U(1)$. This is generated by the Killing vectors $\partial_t$,
$R_i$ and $L_3$ where $L_i$ are the left invariant vector fields on
$SU(2)$ dual to $\sigma_i$ and $R_i$ are the right invariant vector
fields. An interesting question is whether we have enough symmetry
to render geodesic motion in this spacetime Liouville integrable. To
this end we first calculate the Hamiltonian of an uncharged particle
in this background, $H=g^{\mu\nu}p_{\mu}p_{\nu}$. The calculation
amounts to finding the inverse metric $g^{\mu\nu}$. This is most
easily done in an orthonormal frame. We thus define the vielbeins
$\omega^{\mu}$ by: \bea \omega^0 = \frac{r\sqrt{W}}{2b} dt, \qquad
\omega^r = \frac{dr}{\sqrt{W}}, \qquad \omega^1 =
\frac{r}{2}\sigma_1, \qquad  \omega^2 = \frac{r}{2}\sigma_2, \qquad
\omega^3 = b(\sigma_3+fdt) \eea such that $ds^2 =
\eta_{\mu\nu}\omega^{\mu}\omega^{\nu}$. The dual vectors $e_{\mu}$
defined by $ \langle \omega^{\mu} , e_{\nu} \rangle =
\delta^{\mu}_{\nu}$ are: \bea e_0 = \frac{2b}{r\sqrt{W}} \left(
\partial_t - f L_3 \right), \qquad e_r = \sqrt{W} \partial_r \qquad e_1 = \frac{2}{r} L_1, \qquad e_2 =
\frac{2}{r} L_2, \qquad e_3 = \frac{1}{b} L_3. \eea This allows one
to easily write down the inverse metric tensor
$\left(\frac{\partial}{\partial s} \right)^2 = \eta^{\mu\nu} e_{\mu}
e_{\nu} $ which in our case is: \bea \left(\frac{\partial}{\partial
s} \right)^2 = -\frac{4b^2}{r^2 W} \left( \partial_t - fL_3
\right)^2 + W \partial_r^2+\frac{4}{r^2} ( L_1^2 + L_2^2 ) +
\frac{1}{b^2} L_3^2. \eea Define the functions $E=- p_t$ and $M_i =
L_i^{\mu}p_{\mu}$ and $N_i = R_i^{\mu}p_{\mu}$. Then it is clear
that $E$, $M_3$ and $N_3$ must commute with the Hamiltonian $H$ as
they are constants along geodesics. The Hamiltonian for an uncharged
particle is easy to deduce from the inverse metric: \bea H
=-\frac{4b^2}{r^2 W} \left( E + fM_3 \right)^2 + W
p_r^2+\frac{4}{r^2} ( M_1^2 + M_2^2 ) + \frac{1}{b^2} M_3^2. \eea It
is easy to see that $K= M_1^2 +M_2^2 +M_3^2$ Poisson commutes with
the Hamiltonian, as well as $E$, $M_3$ and $N_3$. It is actually the
quadratic Casimir of the $su(2)$ algebra realised by the functions
$M_i$ under the Poisson bracket, $\{ M_i,M_j \}=\epsilon_{ijk}M_k$.
We have thus proved that this Hamiltonian system is Liouville
integrable since we have a five dimensional configuration space and
have found five Poisson commuting functions $H,K,E,M_3,N_3$. In fact
the conserved quantity $K$ corresponds to a reducible Killing tensor
$K_{\mu\nu}$ given by $K=K^{\mu\nu}p_{\mu}p_{\nu}$ which can be read
off as \be K^{\mu\nu} = L_i^{\mu}L_i^{\nu}=R_i^{\mu}R_i^{\nu}.\ee
From the second equality it follows that it is a reducible Killing
tensor. Since the black hole considered here has equal angular momenta
parameters with respect to orthogonal planes, this
result is expected. Presumably a more general charged AdS black hole
would possess an irreducible Killing tensor as has been shown in the
uncharged case~\cite{IntKAdS}. The existence of a Killing tensor is
related to the separability of the Hamilton-Jacobi equation for
particle motion on the space-time. In fact the HJ equation for
particle motion is: \be \frac{\partial S}{\partial l} + g^{\mu\nu}
\frac{\partial S}{\partial
  x^{\mu}} \frac{\partial S}{\partial x^{\nu}}=0
\ee where $S$ is Hamilton's principal function which is a type II
generating function for a canonical transformation $(x^{\mu},
p_{\nu},H) \to (X^{\mu}, P_{\nu},0)$. Thus $p_{\mu} = \partial S/
\partial x^{\mu}$ and $X^{\mu} = \partial S / \partial P_{\mu}$.
Noting \be (L_1 S)^2 + (L_2 S)^2= \left( \cot\theta \partial_{\psi}S
- \frac{1}{\sin\theta} \partial_{\phi}S \right)^2 +
(\partial_{\theta} S)^2 \ee it becomes clear that the
Hamilton-Jacobi equation is separable in the sense \be S = m^2l -Et
+ J_{\psi}\psi + J_{\phi} \phi + \Theta(\theta) +R(r), \ee where
\bea &&\left( \cot\theta J_{\psi} - \frac{1}{\sin\theta} J_{\phi}
\right)^2 + \Theta'(\theta)^2 = C \\ &&-\frac{4b^2(r)}{r^2 W(r)}
\left( E+ f(r)J_{\psi} \right)^2 + W(r)R'(r)^2+\frac{4}{r^2} C +
\frac{1}{b^2} J_{\psi}^2+m^2=0. \eea The constant $C$ is the
separation constant and can easily be related to the conserved
quantity $K$ arising from the Killing tensor $K_{\mu\nu}$ which we
found above. Explicitly $K=C +J_{\psi}^2$. Using the generating
function $S$ one may now obtain the geodesics explicitly, up to
quadratures. By differentiating $S$ with respect to $m^2, C, E,
J_{\psi}, J_{\phi}$ respectively we get: \bea &&l= \frac{1}{2}\int
dr \frac{1}{W R'(r)}, \\ &&\int d\theta \frac{1}{\Theta'(\theta)} =
\int dr \frac{4}{r^2 W} \\ &&t = \int dr
\frac{4b^2(E+fJ_{\psi})}{r^2W^2 R'(r)} \\ &&\psi = \int d\theta
\frac{\cot\theta \left( \cot\theta J_{\psi} -
\frac{J_{\phi}}{\sin\theta} \right)}{\Theta'(\theta)} + \int dr
\left( \frac{ 4b^2f(E+fJ_{\psi})}{r^2W^2 R'(r)} - \frac{J_{\psi}}{b^2WR'(r)} \right) \\
&&\phi = - \int d\theta \frac{ \left( \cot\theta J_{\psi} -
\frac{J_{\phi}}{\sin\theta} \right) }{ \sin\theta \Theta'(\theta)}.
\eea One may also consider the motion of charged particles for which
one needs the Hamiltonian \be
H=g^{\mu\nu}(p_{\mu}+eA_{\mu})(p_{\nu}+eA_{\nu})\ee where $e$ is the
charge of the particle. Since $\mathcal{L}_{V_i} A=0$ for all the
Killing vectors $V_i$ of the black hole spacetime, it immediately
follows that $E,M_3, N_3, K$ still Poisson commute with $H$ and thus
this system is also Liouville integrable.

As a final remark, we should note that the Klein-Gordon equation,
describing a massive scalar field in this background is also
separable as a consequence of the hidden symmetry associated with
the Killing tensor. The general solution looks like $\Phi =
Ne^{-iEt+im_{\psi}\psi+im_{\phi}\phi} P(\cos\theta)f(r)$, where
$P(\cos\theta)$ can be written in terms of Jacobi polynomials.

\section{Thermodynamics}
The black holes studied here can be shown to satisfy the first law
of black hole mechanics~\cite{Ross}, thus suggesting that $S=A/4$.
We would like to make this more concrete by computing the
gravitational partition function and hence the free energy of the
system to check the quantum statistical relation as well as the
first law of thermodynamics. Such a program has been carried out for
the general Kerr-AdS spacetimes in all dimensions in~\cite{GPP}.
\par
We begin by briefly reviewing the thermodynamic quantities for
completeness. The Killing vector $\xi = \partial_t + \Omega_H
2\partial_{\psi}$ is null on the horizon where \be \Omega_H =
-f(r_+)/2. \ee The potential measured in a co-rotating frame on the
horizon, defined by $\phi_H = \xi^{\mu}A_{\mu}$, is \be \Phi_H =
\frac{\sqrt{3}q}{2r_+^2}( 1-j\Omega_H). \ee The area of the horizon
is easily computed to be \be A= 4\pi^2r_+^2 b(r_+) \ee as is the
surface gravity $\kappa$ which we find to be \be \kappa = \frac{r_+
W'(r_+)}{4 b(r_+)}. \ee Armed with these quantities one may verify
that \be \label{firstlaw} dM = \frac{\kappa}{8\pi} dA + 2\Omega_H dJ
+ \Phi_H dQ \ee which defines the thermodynamic mass $M$ to be \be
M= \frac{\pi}{4}( 3p-3q- \lambda j^2 p). \ee This is a tedious task
for which one needs to use the fact that $W(r_+)=0$. We should pause
to emphasise the non-triviality of this statement. We can only
define such a thermodynamic mass because the RHS of (\ref{firstlaw})
is an exact differential. Also, we should emphasise that
$M=Q[\partial_t]$ which is the statement that the mass~(\ref{ADMmass})
computed from the prescription of Ashtekar, Magnon, and
Das agrees with the thermodynamic mass.
\par
To compute the partition function we will follow the standard
prescription~\cite{GH} and carry out the path integral
along a complex contour and then analytically continue back.
\par \noindent
A real Euclidean section of the charged, rotating, asymptotically
AdS black hole can be found upon making the substitutions $t = -i
\tau$ and $j = i \hat{j}$. Note that the charge parameter $q$ cannot
be so analytically continued and so the Maxwell field is pure
imaginary on the Euclidean section. For the following computation we
will henceforth work with this Euclidean section. Let the largest positive root
of $W(r)$ be $r_+$. Removal of the conical singularity at $r=r_+$
requires that $\tau$ be identified with period
\begin{equation}
\beta = \frac{1}{T} = \frac{8\pi b(r_{+})}{r_{+}W'(r+)}
\end{equation}
thus confirming the standard relation $T=\kappa/2\pi$.
\par \noindent
The on-shell Euclidean action is
\begin{equation}
\label{onshell} \hat{I} = -\frac{1}{2\pi} \int_{M} \sqrt{g} \lambda
+ \frac{1}{12\pi}\int_{\partial M} \sqrt{h}F^{\mu \nu}A_{\nu}n_{\mu}
- \frac{1}{8\pi} \int_{\partial M} \sqrt{h} K
\end{equation} where we have included the usual Gibbons-Hawking boundary term.
The Euclidean action $\hat{I}$ is defined via $I = i\hat{I}$. Here
the boundary of the spacetime $M$ is taken to be a hypersurface $r =
\const$ with the limit $r \rightarrow \infty$ taken after performing
the integral. The bulk term is obviously divergent, as in the case
for asymptotically AdS spacetimes.
\par \noindent
We begin with the second term, the contribution from the Maxwell
field. Firstly, since the Killing vector $\xi$ has zero norm on the
horizon we require $\xi \cdot A=0$ on the horizon. This is achieved
upon performing the  gauge transformation $A \rightarrow A - \Phi_H
dt \equiv \tilde{A}$. It is easiest to work in the vielbein basis.
Explicitly
\begin{equation}
\tilde{A} =
\frac{2b}{r\sqrt{W}}\frac{\sqrt{3}q}{2r^2}\left(1+\frac{fj}{2}
\right )\omega^{0} - \frac{\sqrt{3}qj}{4br^2} \omega^{3} -A_H
\end{equation} where we write $A_H = \Phi_H dt$. Note that $\sqrt{h}=\frac{\sqrt{W}r^3\sin{\theta}}{8} $. Now, since
$n^{\mu} = e_r^{\mu}$, it is clear that the only relevant components
of the field strength are $F_{\hat{r} \hat{0}}$ and $F_{\hat{r}
\hat{3}}$. It is easy to compute $F$ and we find: \be F=
-\frac{2\sqrt{W}}{r} \omega^r \wedge A - \frac{\sqrt{3}qj}{4r^2}
\sigma_1 \wedge \sigma_2 \ee from which we extract $F_{\hat{r}
\hat{0}}$ and $F_{\hat{r} \hat{3}}$. By studying the asymptotics
(for large $r$) of these components of $F$, $A$ and $\sqrt{h}$ one
may easily see that the only non-zero contribution to the quantity
$\sqrt{h}F_{\mu\nu}\tilde{A}^{\nu}n^{\mu}$ is from
$-\sqrt{h}F_{\hat{r} \hat{0}} A_H^{\hat{0}}$. We get \bea \sqrt{h}
F_{\mu\nu}\tilde{A}^{\nu}n^{\mu} \sim -\frac{3}{16}\sin\theta
\frac{q^2}{r_+^2}(1-j\Omega_H) \eea which allows us to get \be
\frac{1}{12\pi}\int_{\partial M}\sqrt{h}
F_{\mu\nu}\tilde{A}^{\nu}n^{\mu} =  -\frac{\pi \beta
q^2}{4r_+^2}(1-j\Omega_H). \ee
\par \noindent
Now we turn to the calculation of the first term in (\ref{onshell}),
which is somewhat more subtle. The reason is that as it stands it is
divergent due to AdS having infinite volume, and hence we need to
regularize it. The method consists of performing the integral over
the $r$ coordinate only up to some finite, yet large, value which we
shall denote by $R$. Then we subtract the same integral evaluated
for the background AdS metric which we may get from the black hole
by setting $p=q=0$. However there is a subtlety. The black hole
metric and the metric for AdS must match at $r=R$ for large $R$.
This can be taken care of by rescaling the $\tau$ coordinate such
that: \bea (1-\lambda R^2)\tau_0^2 = W(R) \tau^2. \eea This can be
thought of as a rescaling of the temperature \be \beta_0 = \sqrt{
\frac{W(R)}{1-\lambda R^2}} \beta. \ee The final piece of
information we need is $\sqrt{g} = r^3\sin\theta /8$. It is then
easy to calculate the integrals \bea \int_{r_+ \leq r \leq R}
\sqrt{g} - \int_{0 \leq r \leq R} \sqrt{\bar{g}} = \frac{\pi^2}{2}
(R^4-r_+^4) (\beta-\beta_0) -\frac{\pi^2}{2}r_+^4\beta_0 \eea where
$\bar{g}$ is the $AdS$ metric with the $\tau$ coordinate rescaled as
discussed above. Next we need to find the limit of this expression
as $R \to \infty$. To do this note that \be
\frac{\beta-\beta_0}{\beta} = -\frac{p(1+\lambda j^2)-q}{\lambda
R^4} + O(R^{-6}) \ee which allows one to deduce that \bea
-\frac{\lambda}{2 \pi} \left( \int_{r_+ \leq r \leq R} \sqrt{g} -
\int_{0 \leq r \leq R} \sqrt{\bar{g}} \right) \to \frac{\pi \beta
(p(1+\lambda j^2)-q)}{4} + \frac{\pi}{4} \lambda r_+^4 \beta \eea as
$R \to \infty$. This is the regularised contribution to the action
from the bulk integral.
\par \noindent Finally we turn to the last surface term involving
the extrinsic curvature. This is facilitated by using the identity
$\mathcal{L}_n \sqrt{h} = \sqrt{h} K$. This integral will also have
to be regularised by the subtraction method used for the bulk
integral. It is then straightforward to show \bea \int_{r=R}
\sqrt{h} - \int_{r=R}\sqrt{\bar{h}} = \const + O(R^{-2}) \eea \ and
thus \bea \int_{r=R} \sqrt{h}K -\int_{r=R} \sqrt{\bar{h}}\bar{K} =
\sqrt{W(R)}\frac{\partial}{\partial R}\left(\int_{r=R} \sqrt{h} -
\int_{r=R} \sqrt{\bar{h}} \right) = O(R^{-2}) \eea Hence, sending $R
\to \infty$ shows that the contribution from the Gibbons-Hawking
term to the gravitational action vanishes in our case, just like in
the uncharged case. Collecting all these results gives us the
Euclidean action for the black hole \be \hat{I} =\frac{\pi \beta
(p(1+\lambda j^2)-q)}{4} -\frac{\pi \beta q^2}{4r_+^2}(1-j\Omega_H)
+\frac{\pi}{4}\lambda r_+^4\beta. \ee Note that in the uncharged
case ($q=0$) this agrees with the results of~\cite{HHT,GPP} when the
rotation parameters are equal. To see this one needs to know that
$p=m/\Xi^3$, $j=a$ and $r^2 \to (r^2+a^2)/\Xi$ with of course
$\lambda =-1/l^2$ and $\Xi =1-a^2/l^2$.
\par
We now turn to the verification of the quantum statistical
relation~\cite{GPP}. In a thermal ensemble of fixed temperature,
angular velocity, and potential, the partition function satisfies $Z
= e^{-\beta \Phi}$ where $\Phi$ is the Gibbs free energy. From the
Euclidean quantum gravity point of view, in the semiclassical limit
we expect
\begin{equation}\label{QSR}
\beta \Phi = \hat{I}
\end{equation} to hold. As first observed in~\cite{GH} this relation
involves Planck's constant and is quantum mechanical in origin.
Using the fact that $W(r_+)=0$, it is a simple matter of tedious algebra to check that \be \label{qsr}
T\hat{I} = M-TS -2\Omega_H J -\Phi_H Q \ee if and only if $S=A/4$. This can be
thought of in two ways. Either a proof of the fact that $S=A/4$ for
this black hole given~(\ref{QSR}) or alternatively one may
suppose $S=A/4$ and deduce the quantum statistical relation
(\ref{qsr}) above. The former viewpoint seems to be
the more logical standpoint. In fact ideally one should express the
partition function in terms of the intensive variables $(\beta,
\Omega_H, \Phi_H)$, from which one can then derive the extensive
quantities $(S,J,Q)$. This appears to be rather difficult.
\par
When $\lambda=0$ one may verify the Smarr relation
\be
M= \frac{3}{2} TS + \frac{3}{2} 2\Omega_H J +\Phi_H Q
\ee
using the fact that $W(r_+)=0$. When $\lambda \neq 0$ this relation
does not hold. This can be thought of as due to having an extra
dimensionful quantity, $\lambda$, which breaks the scale invariance of
the space-time. In fact, if one treats $\lambda$ as one of the
extensive variables of our thermodynamic system, one can write down a
Smarr relation:
\be
M=\frac{3}{2} TS + \frac{3}{2} 2\Omega_H J +\Phi_H Q - \Theta \lambda
\ee
where the intensive variable $\Theta$ conjugate to $\lambda$ is given
by:
\bea
\Theta = -\frac{\pi}{32 r_+^4 b(r_+)^2} (3r_+^{10} +10pj^2r_+^6 +8p^2j^4 r_+^2-3j^2q^2r_+^4 -4j^4q^2p).
\eea
From this one may now deduce a generalised first law using Euler's
theorem for homogeneous functions. To do this we note that
$M=M(S,Q,J,\lambda)$ and that the dimensions, in units of length are
$[M]=2$, $[Q]=2$, $[J]=3$, $[S]=3$, $[\lambda]= -2$. Then application of
Euler's theorem tells us
\be
M= \frac{3}{2} \frac{\partial M}{\partial S} S + \frac{3}{2}\frac{\partial
  M}{\partial J} J + \frac{\partial M}{\partial Q} Q - \frac{\partial
  M}{\partial \lambda} \lambda. \ee
From this it follows that:
\bea
dM = T dS + 2\Omega_H dJ + \Phi_H dQ + \Theta d\lambda.
\eea
\section{Discussion} We have been primarily concerned with
properties of the solution from the five dimensional standpoint. We
conclude with some remarks on some applications.
Solutions of minimal $N=1$ gauged supergravity may be uplifted into
Type IIB supergravity by a straightforward oxidation method. One can
hence construct a general non-supersymmetric configuration from the
black holes studied here. More precisely if $ds_{5}^2$ represents
the five dimensional metric~(\ref{solution}) then the
ten-dimensional solution reads
\begin{equation}
ds^2_{10} = ds_{5}^2 + \ell^2 \sum_{i=1}^{3} \left( d\mu_{i}^2 +
  \mu^2_{i}\left(d\xi_{i} + \frac{2}{\ell \sqrt{3}}A \right )^2 \right).
\end{equation} Here $\sum_{i}^{3}\mu_{i}^2 = 1$ and $0 \leq \xi_{i} <
  2\pi$ parameterise $S^{5}$. The RR five form flux is
\begin{equation}
 F_{[5]} = ( 1 + \star_{10})\left ( -\frac{4}{\ell}\rm{vol}_{5}
 + \frac{\ell^2}{\sqrt{3}} \sum_{i=1}^{3} d(\mu_{i}^2) \wedge d\xi_{i}
 \wedge \star_{5}F_{[2]} \right ).
\end{equation} Here $\rm{vol}_{5}$ is the volume form
 on~(\ref{solution}). These solutions are simply non extremal generalisations
 of the supersymmetric solutions preserving 2 supersymmetries found
 in~\cite{GGS}.

From the string point of view, it is natural to consider these charged,
 rotating black holes solutions in the context of the $AdS_5/CFT_4$
 correspondence. Alternatively, one can take holography at face value
 and try to match properties of the bulk spacetime with a particular
 conformal field theory living on its boundary. This has been done in
 the singly rotating case~\cite{Berman} and the static charged case~\cite{CEJM} in the high
 temperature limit.  It is in this limit that the supergravity action,
 corresponding to the strongly coupled region of the gauge theory, is
 found to match to the weakly coupled field theory action evaluated on
 the boundary up to the usual $\frac{4}{3}$ factor. Other limits
 that have been explored include $\Omega \to 1$~\cite{HHT}.

In this charged rotating case one expects similar results. As
usual take $\cal{N}$$=4$, $SU(N)$ super Yang-Mills on
 $S^3$ of radius $R$. Note the boundary of the black hole metric is
 \emph{not} rotating. The `charge' on the
 field theory side originates from the various fields carrying equal $R$-charges
 in the
three $U(1)$ Cartan subgroups of the $SO(6)$ (see~\cite{Obers} for a
more general treatment). One can show in the high temperature limit
that the contribution to the partition function from the charge is
sub-leading with respect to the angular momentum. Explicitly we have
\bea F_{CFT}(T,\Omega,\Phi) &&= T_{CFT} \sum_i \int_0^{\infty}dl_i
\int_{-l_i}^{l_i} dm_L^i \int_{-l_i}^{l_i} dm_R^i \sum_{q_i}
\log(1-\eta_i e^{-\beta(\l_i/R +2\Omega m_L^i + q_i \Phi)}) \nonumber \\
&&\sim -\frac{\pi^2}{6}\frac{N^2 V T^4_{CFT}}{(4R^2\Omega^2 -1)^2} \eea where the
sum over $i$ denotes summing over the different fields in the CFT,
$\eta_i$ is $+1$($-1$) for a boson (fermion), $m_L$, $m_R$ and $l$
are the quantum numbers of the $SU(2)_L\times SU(2)_R$ symmetry and
$q$ the quantum numbers of the $U(1)$ $R$-charge. On the strongly
coupled side, it is more subtle to see how the high
 temperature limit should be taken, given the difficulty in writing the
 action in terms of the intensive variables. It would be interesting
 to study this limit, as well as the thermodynamic stability of these
 solutions.

Finally, one could extend the results of this paper to the $N=2$, $U(1)^3$, gauged
supergravity where similar non-extremal black holes were found~\cite{Cvetic}. \\

\noindent {\bf \large Acknowledgments:} The authors would like to
thank David Berman for discussions on the holography. HKK would like
to thank St.\ John's College, Cambridge, for financial support.

\appendix
\section{Useful formulae}
First we define the various functions appearing in the metric.
Firstly we have \be W(r) = 1- 4\lambda b(r)^2 - \frac{1}{r^2}(2p-2q)
+\frac{1}{r^4}(q^2 + 2pj^2) \ee where the function $b(r)$ is given
by \be b(r)^2 = \frac{r^2}{4} \left( 1- \frac{j^2q^2}{r^6} +
\frac{2j^2p}{r^4} \right). \ee The function $f(r)$ is given by \be
f(r) = -\frac{j}{2b(r)^2} \left( \frac{2p-q}{r^2} -\frac{q^2}{r^4}
\right). \ee We have also made use of the left invariant 1-forms on
$SU(2)$, denoted by $\sigma_i$ where $i=1,2,3$. They are defined by:
\bea &&\sigma_1=-\sin\psi d\theta +\cos\psi\sin\theta d\phi \\ &&\sigma_2= \cos\psi d\theta +\sin\psi\sin\theta d\phi \\
&&\sigma_3 = d\psi +\cos\theta d\phi \eea where $0\leq \theta \leq
\pi$, $0 \leq \phi \leq 2\pi$ and $0\leq \psi \leq 4\pi$. The left
invariant vector fields $L_i$ which form a dual basis to $\sigma_i$,
so $\langle \sigma_i , L_j \rangle =\delta_{ij}$ are:
\bea &&L_1= -\cot\theta \cos\psi \partial_{\psi} -\sin\psi\partial_{\theta} +\frac{\cos\psi}{\sin\theta} \partial_{\phi} \\
&&L_2 = -\cot\theta \sin\psi \partial_{\psi} +\cos\psi\partial_{\theta} +\frac{\sin\psi}{\sin\theta} \partial_{\phi} \\
&&L_3 =\partial_{\psi}. \eea The right invariant vector fields are
denoted by $R_i$, and are given by:
\bea &&R_1= \cot\theta \cos\phi \partial_{\phi} +\sin\phi\partial_{\theta} -\frac{\cos\phi}{\sin\theta} \partial_{\psi} \\
&&R_2 = -\cot\theta \sin\phi \partial_{\phi} +\cos\phi\partial_{\theta} +\frac{\sin\phi}{\sin\theta} \partial_{\psi} \\
&&R_3 =\partial_{\phi}. \eea

\end{document}